\documentclass[11pt,a4paper]{article}

\usepackage{epsfig,amsmath,amssymb,cite}
\usepackage{hyperref}

\begin{document}

\title{Thin shells in $F(R)$ gravity with non-constant scalar curvature} 
\author{Ernesto F. Eiroa\thanks{e-mail: eiroa@iafe.uba.ar} , Griselda Figueroa Aguirre\thanks{e-mail: gfigueroa@iafe.uba.ar}\\
{\small  Instituto de Astronom\'{\i}a y F\'{\i}sica del Espacio (IAFE, CONICET-UBA),}\\
{\small Casilla de Correo 67, Sucursal 28, 1428, Buenos Aires, Argentina}} 
\date{}
\maketitle

\begin{abstract}
We introduce two classes of spherically symmetric spacetimes having a thin shell of matter, in non-quadratic $F(R)$ theories of gravity with non-constant scalar curvature $R$. In the first, the thin shell joins an inner region with an outer one, while in the second it corresponds to the throat of a wormhole. In both scenarios, we analyze the stability of the static configurations under radial perturbations. As particular examples of spacetimes with a cosmological constant, we present charged thin shells surrounding a non-charged black hole and charged thin-shell wormholes. We show that in both cases stable solutions are possible for suitable values of the parameters.
\end{abstract}

\section{Introduction}\label{intro} 

The difficulties of general relativity in providing a fully accepted explanation to some problems, such as singularities, the nature of the dark matter, the understanding of the accelerated expansion of the Universe, or the case of quantum gravity, motivate the study of modified gravity theories. The simplest proposal is the so-called $F(R)$ theories \cite{dft,sofa,nojod}, which generalize the Einstein-Hilbert action by replacing the Ricci scalar curvature $R$ with a function of it. They have also an equivalence with the scalar-tensor gravity. These theories offer important applications such as the Starobinsky model which explains the early inflation era of the Universe, and they can give a possible explanation of the acceleration of the Universe expansion without using dark energy. They also provide interesting models for topics such as the cosmic microwave background, weak lensing, structure formation, and the spectra of galaxy clustering, among others. 
There exist many static spherically symmetric solutions of the field equations in $F(R)$ gravity that can be gathered into two groups: one with constant scalar curvature and the other with non-constant scalar curvature. The first group contains solutions that describe branes \cite{branefr}, traversable wormholes \cite{whfr}, and black holes \cite{sofa,bhfr1,bhfr2,nojod,bhfr3,bhfr4}. The second one consists mostly of black hole solutions \cite{rnonconst1,rnonconst2,rnonconst3,rnonconst4,rnonconst5}. Neutron stars \cite{ns-fr} and domain wall solutions \cite{dw-fr} are also topics of interest addressed in the literature.
 
The conditions for the proper construction of thin shells of matter by cutting and pasting two solutions in general relativity is provided by the well-known Darmois-Israel formalism \cite{daris}. It allows to analyze the characteristics and the dynamics of a thin layer of matter at the matching hypersurface. Due to its simplicity and its malleability, this tool has been used to study models of gravastars \cite{gravstar1,gravstar2}, vacuum bubbles and thin layers around black holes \cite{sh1,sh2,sh3}, and wormholes \cite{wh1,wh2,wh3,whcil}. The symmetric nature, spherical or cylindrical, of the constructions in these works allows a much easier analysis of their stability under radial perturbations. In comparison with general relativity, the junction formalism in $F(R)$ theories \cite{dss,js1} becomes more restrictive since it introduces a constraint on the trace of the second fundamental form, which has to be continuous at the joining hypersurface \cite{js1}. For non-quadratic $F(R)$, the continuity of the scalar curvature at both sides of it is added as an extra condition. However, for the quadratic case, this last condition can be relaxed, and as a consequence, new contributions to the standard energy--momentum tensor appear, consisting of an external scalar pressure/tension, an external energy flux vector, and a double layer energy--momentum tensor which resembles classical dipole distributions \cite{js1,js2-3,js4}. The junction formalism in $F(R)$ gravity has been used in models of thin-shell wormholes \cite{eirfig,whfrts,whfrtsnew}, bubbles \cite{bb1}, and thin shells of matter surrounding black holes \cite{bb1}. 

In the current work, we construct spherically symmetric thin shells in non-quadratic $F(R)$ gravity with non-constant scalar curvature by using the corresponding junction conditions and we analyze the stability of the static solutions under perturbations preserving the symmetry. We study two configurations: an inner geometry joined across the shell to an outer one and a thin-shell wormhole. We consider an example for each situation, one consisting of a charged thin shell of matter enclosing a non-charged black hole and the other of a charged thin-shell wormhole. In both cases, we adopt the particular gravitational Lagrangian $F(R)=R+2\beta\sqrt{R-8\Lambda}-\Lambda$, with $\beta <0$ and $\Lambda$ twice the conventional cosmological constant. In Sects. \ref{tsh} and \ref{whs}, we introduce the two general formalisms, while in Sect. \ref{examples} we provide the specific examples. Finally, in Sect. \ref{discu} we present a discussion of the results. We use units so that $c=G=1$, where $c$ is the speed of light and $G$ the gravitational constant.

\section{Construction and stability of thin shells}\label{tsh}

We proceed to construct a manifold $\mathcal{M}$ composed of the union of two different ones, $\mathcal{M}_1$ and $\mathcal{M}_2$, using the junction conditions in $F(R)$ gravity. The matching of these two manifolds is done at a hypersurface $\Sigma$ consisting of a thin shell of matter. We start with two spherically symmetric spacetimes, with the metrics
\begin{equation}
ds^2=-A_{1,2} (r) dt_{1,2}^2+A_{1,2} (r)^{-1} dr^2+r^2(d\theta^2 + \sin^2\theta d\varphi^2),
\label{sphericalmetric}
\end{equation}
where $t_{1,2}$ is the corresponding time coordinate, $r>0$ is the radial coordinate, and $0\le \theta \le \pi$ and $0\le \varphi<2\pi $ are the angular coordinates. We work with a scalar curvature that may depend on $r$, i.e., $R_{1,2}=R_{1,2}(r)$, then for the metrics above their expressions take the form
\begin{equation}
R_{1,2}(r)= -A_{1,2}''(r)-\frac{4}{r}A_{1,2}'(r)-\frac{2}{r^2}A_{1,2}(r)+\frac{2}{r^2},
\label{Rnonconstant}
\end{equation}
where the prime on $A_{1,2}(r)$ means the derivative with respect to $r$, while the corresponding derivatives read
\begin{equation}
R_{1,2}'(r)= -A_{1,2}'''(r)-\frac{4}{r}A_{1,2}''(r)+\frac{2}{r^2}A_{1,2}'(r)+\frac{4}{r^3}A_{1,2}(r)-\frac{4}{r^3},
\label{DerRnonconstant}
\end{equation}
which will be used in further calculations. We identify $\mathcal{M}_1$ as the inner zone given by $0\leq r \leq a$ and $\mathcal{M}_2$ as the outer region defined by $r \geq a$. When we match them at the radius $a$, the new manifold $\mathcal{M}=\mathcal{M}_1 \cup \mathcal{M}_2$ is obtained, where a global radial coordinate  $r\in [0,+\infty)$ is defined and the angular coordinates are mutually identified. The complete spacetime $\mathcal{M}$ is described by the coordinates $X^{\alpha }_{1,2} = (t_{1,2},r,\theta, \varphi)$. On the matching hypersurface $\Sigma$, corresponding to $G(r)\equiv r-a =0$, we choose the coordinates $\xi ^{i}=(\tau ,\theta,\varphi )$, with $\tau $ the proper time. We let the radius of this hypersurface depend on $\tau $, i.e., $a(\tau)$, and we denote its derivative with respect to $\tau$ by $\dot{a}(\tau)$. The proper time at the sides of the shell should be the same, so that
\begin{equation*}
\frac{dt_{1,2}}{d\tau} = \frac{\sqrt{A_{1,2}(a) + \dot{a} ^2}}{A_{1,2}(a)},
\end{equation*}
in which the free signs were fixed by demanding the times $t_{1,2}$ and $\tau$ all run into the future.

The first fundamental form at the sides of the shell is given by
\begin{equation}
h^{1,2}_{ij}= \left. g^{1,2}_{\mu\nu}\frac{\partial X^{\mu}_{1,2}}{\partial\xi^{i}}\frac{\partial X^{\nu}_{1,2}}{\partial\xi^{j}}\right| _{\Sigma },
\end{equation}
and the second fundamental form reads
\begin{equation}
K_{ij}^{1,2 }=-n_{\gamma }^{1,2 }\left. \left( \frac{\partial ^{2}X^{\gamma
}_{1,2} } {\partial \xi ^{i}\partial \xi ^{j}}+\Gamma _{\alpha \beta }^{\gamma }
\frac{ \partial X^{\alpha }_{1,2}}{\partial \xi ^{i}}\frac{\partial X^{\beta }_{1,2}}{
\partial \xi ^{j}}\right) \right| _{\Sigma },
\label{sff}
\end{equation}
where the unit normals ($n^{\gamma }n_{\gamma }=1$) are determined by
\begin{equation}
\label{general_normal_fr}
n_{\gamma }^{1,2 }=\left\{ \left. \left| g^{\alpha \beta }_{1,2}\frac{\partial G}{\partial
X^{\alpha }_{1,2}}\frac{\partial G}{\partial X^{\beta }_{1,2}}\right| ^{-1/2}
\frac{\partial G}{\partial X^{\gamma }_{1,2}} \right\} \right| _{\Sigma },
\end{equation}
and they are taken to point from $\mathcal{M}_1$ to $\mathcal{M}_2$. On the surface $\Sigma $, we choose to work in the orthonormal basis given by $\{ e_{\hat{\tau}}=e_{\tau }, e_{\hat{\theta}}=a^{-1}e_{\theta }, e_{\hat{\varphi}}=(a\sin \theta )^{-1} e_{\varphi }\} $. Therefore, and taking into account the metrics (\ref{sphericalmetric}), we obtain that the first fundamental form is $h^{1,2}_{\hat{\imath}\hat{\jmath}}= \mathrm{diag}(-1,1,1)$, the unit normals result in
\begin{equation}
n_{\gamma }^{1,2}= \left(-\dot{a},\frac{\sqrt{A_{1,2}(a)+\dot{a}^2}}{A_{1,2}(a)},0,0 \right),
\label{normal:dynamic}
\end{equation}
and the non-null components of the second fundamental form are
\begin{equation}
K_{\hat{\tau}\hat{\tau}}^{1,2 }=-\frac{A '_{1,2}(a)+2\ddot{a}}{2\sqrt{A_{1,2}(a)+\dot{a}^2}}
\label{Ks:dynamic1}
\end{equation}
and
\begin{equation}
K_{\hat{\theta}\hat{\theta}}^{1,2}=K_{\hat{\varphi}\hat{\varphi}}^{1,2
}=\frac{1}{a}\sqrt{A_{1,2} (a) +\dot{a}^2}.
\label{Ks:dynamic2}
\end{equation}
The jump of any quantity $\Upsilon $ across the hypersurface $\Sigma$ is defined by $[\Upsilon ]\equiv (\Upsilon ^{2}-\Upsilon  ^{1})|_\Sigma $. The junction conditions in $F(R)$ gravity \cite{js1} require the continuity of the first fundamental form so that $[h_{\mu \nu}]=0$, which is satisfied automatically in this case, and the continuity of the trace of the second fundamental form, 
\begin{equation}
[K^{\mu}_{\;\; \mu}]=0
\label{conditionEq}
\end{equation}
that, by using Eqs. (\ref{Ks:dynamic1}) and (\ref{Ks:dynamic2}), leads to
\begin{equation}
-\frac{2a\ddot{a}+a A_{1}'(a) + 4(A_{1}(a)+\dot{a}^2)}{\sqrt{A_{1}(a)+\dot{a}^2}}+\frac{2a\ddot{a}+a A_{2}'(a) + 4(A_{2}(a)+\dot{a}^2)}{\sqrt{A_{2}(a)+\dot{a}^2}}=0,
\label{KJump:dynamic01}
\end{equation}
or as well
\begin{equation}
\frac{2\ddot{a}+ A_{2}'(a)}{2\sqrt{A_{2}(a)+\dot{a}^2}}-\frac{2\ddot{a}+ A_{1}'(a)}{2\sqrt{A_{1}(a)+\dot{a}^2}}=\frac{2}{a}\left(\sqrt{A_{1}(a)+\dot{a}^2}-\sqrt{A_{2}(a)+\dot{a}^2}\right)\label{KJump:dynamic02}.
\end{equation}
Additionally, in non-quadratic $F(R)$, i.e., when $F'''(R) \neq 0$ (here the prime denotes the derivative with respect to $R$), the formalism requires the continuity of the scalar curvature across the hypersurface \cite{js1} 
\begin{equation}
[R]=0,
\label{R_same}
\end{equation}
so that $R_1(a)=R_2(a) \equiv R_\Sigma $. The field equations at the shell $\Sigma $ have the form \cite{js1} 
\begin{equation}
\kappa  S_{\mu \nu}=-F'(R_\Sigma)[K_{\mu \nu}]+ F''(R_\Sigma)[\eta^\gamma \nabla_\gamma R]  h_{\mu \nu}, \;\;\;\; n^{\mu}S_{\mu\nu}=0,
\label{fieldeq}
\end{equation}
with $\kappa =8\pi $, $S_{\mu \nu}$ the surface energy--momentum tensor, and $\nabla_\gamma $ the covariant derivative. In quadratic $F(R)$, i.e., when $F'''(R) = 0$, the scalar curvature can be discontinuous and the field equations have a different form at the shell  \cite{js1}. In this case, the extra contributions to the energy--momentum tensor mentioned in Sect. \ref{intro} appear; for further details on this matter, see Ref. \cite{js1}. In what follows, we are only interested in the study of thin shells in the non-quadratic theory. In the orthonormal basis, the energy--momentum tensor takes the form  $S_{_{\hat{\imath}\hat{\jmath} }}={\rm diag}(\sigma ,p_{\hat{\theta}},p_{\hat{\varphi}})$, with $\sigma$ the surface energy density and $p_{\hat{\theta}}=p_{\hat{\varphi}}=p$ the transverse pressures. Using Eq. (\ref{fieldeq}), we obtain
\begin{eqnarray}
\sigma &=& \frac{F'(R_\Sigma)}{2\kappa }\left(  \frac{2\ddot{a}+A_{2}'(a)}{\sqrt{A_{2}(a)+\dot{a}^2}}- \frac{2\ddot{a}+A_{1}'(a)}{\sqrt{A_{1}(a)+\dot{a}^2}}\right) \nonumber \\
&&-\frac{F''(R_\Sigma)}{\kappa }\left( R_2'(a)\sqrt{A_{2}(a)+\dot{a}^2}- R_1'(a)\sqrt{A_{1}(a)+\dot{a}^2}\right) 
\label{sigma:dynamic}
\end{eqnarray}
and
\begin{eqnarray}
p&=&-\frac{F'(R_\Sigma)}{\kappa a}\left(  \sqrt{A_{2}(a)+\dot{a}^2}-  \sqrt{A_{1}(a)+\dot{a}^2}\right) \nonumber \\
&&+\frac{F''(R_\Sigma)}{\kappa }\left( R_2'(a)\sqrt{A_{2}(a)+\dot{a}^2}- R_1'(a)\sqrt{A_{1}(a)+\dot{a}^2}\right),
\label{pressure:dynamic}
\end{eqnarray}
where the scalar curvature and its derivative are given by Eqs. (\ref{Rnonconstant}) and (\ref{DerRnonconstant}). The constraint $F'(R)>0$ guarantees to have a positive effective Newton constant $G_{\mathrm{eff}} = G/F'(R) = 1/F'(R)$, preventing the graviton to be a ghost \cite{nojod}; further discussion can be found in Ref. \cite{bronnikov}. The weak energy condition, given in the orthonormal basis by $\sigma \geq 0$ and $\sigma + p \geq 0$, determines the kind of matter at the shell: when it is satisfied, the matter is normal; otherwise, it is dubbed exotic. Using Eq. (\ref{KJump:dynamic02}), one can see that Eq. (\ref{sigma:dynamic}) can be rewritten in the form 
\begin{eqnarray}
\sigma &=& -\frac{2 F'(R_\Sigma)}{\kappa a}\left( \sqrt{A_{2}(a)+\dot{a}^2}-\sqrt{A_{1}(a)+\dot{a}^2}\right) \nonumber \\
&& -\frac{F''(R_\Sigma)}{\kappa }\left( R_2'(a)\sqrt{A_{2}(a)+\dot{a}^2}- R_1'(a)\sqrt{A_{1}(a)+\dot{a}^2}\right);
\label{sigma:dynamic_plus_cond}
\end{eqnarray}
from this equation and using Eq. (\ref{pressure:dynamic}), the equation of state is 
\begin{equation}
\sigma -2p=-3\lambda   
\label{state_eq:dynamic}
\end{equation}
with
\begin{equation}
\lambda= \frac{F''(R_\Sigma)}{\kappa }\left( R_2'(a)\sqrt{A_{2}(a)+\dot{a}^2}- R_1'(a)\sqrt{A_{1}(a)+\dot{a}^2}\right) 
\label{BraneTension:dynamic}
\end{equation} 
the brane tension of the matching hypersurface \cite{js1}. 

In the case of static configurations with a constant radius $a_0$, it is straightforward that Eq. (\ref{conditionEq}) becomes 
\begin{equation}
-\frac{a_0 A_{1}'(a_0) + 4A_{1}(a_0)}{\sqrt{A_{1}(a_0)}}+\frac{a_0 A_{2}'(a_0)+ 4A_{2}(a_0)}{\sqrt{A_{2}(a_0)}}=0
\label{KJump:static01}
\end{equation}
or
\begin{equation}
\frac{A_{2}'(a_0)}{2\sqrt{A_{2}(a_0)}}-\frac{A_{1}'(a_0)}{2\sqrt{A_{1}(a_0)}}=\frac{2}{a_0}\left(\sqrt{A_{1}(a_0)}-\sqrt{A_{2}(a_0)}\right),
\label{KJump:static02}
\end{equation}
while the energy density and the pressure at $\Sigma $ read
\begin{equation} 
\sigma_0 = -\frac{2 F'(R_\Sigma)}{\kappa a_0}\left( \sqrt{A_{2}(a_0)}-  \sqrt{A_{1}(a_0)}\right)
-\frac{F''(R_\Sigma)}{\kappa }\left( R_2'(a_0)\sqrt{A_{2}(a_0)}- R_1'(a_0)\sqrt{A_{1}(a_0)}\right) 
\label{sigma:static_plus_cond}
\end{equation}
and
\begin{equation}
p_0 = -\frac{F'(R_\Sigma)}{\kappa a_0}\left( \sqrt{A_{2}(a_0)}- \sqrt{A_{1}(a_0)} \right) 
+\frac{F''(R_\Sigma)}{\kappa }\left( R_2'(a_0)\sqrt{A_{2}(a_0)}- R_1'(a_0)\sqrt{A_{1}(a_0)}\right) ,
 \label{pressure:static_plus_cond}
\end{equation}
which fulfill the equation of state
\begin{equation}
\sigma_0 -2p_0=-3\lambda _0  
\label{state_eq:static}
\end{equation}
where
\begin{eqnarray}
\lambda_0 &=& \frac{F''(R_\Sigma)}{\kappa }\left( R_2'(a_0)\sqrt{A_{2}(a_0)}-  R_1'(a_0)\sqrt{A_{1}(a_0)}\right) 
\label{BraneTension:static}
\end{eqnarray}
is the brane tension in this case.

In order to analyze the stability of the solutions under radial perturbations, we rewrite Eq. (\ref{KJump:dynamic01}) using $\ddot{a}= (1/2)d(\dot{a}^2)/da$ and the definition of $z=\sqrt{A_{2}(a)+\dot{a}^2}-\sqrt{A_{1}(a)+\dot{a}^2}$, so we have to solve the equivalent equation $az'(a)+2z(a)=0$. Its solution gives us an expression for $\dot{a}^{2}$ in terms of an effective potential
\begin{equation}
\dot{a}^{2}=-V(a),
\label{condicionPot}
\end{equation}
where 
\begin{equation}
V(a)= -\frac{a_{0}^4\left(\sqrt{A_{2}(a_{0})}-\sqrt{A_{1}(a_{0})}\right)^2}{4a^4} +\frac{A_{1}(a)+A_{2}(a)}{2} -\frac{a^4 \left(A_{2}(a)-A_{1}(a)\right)^{2}}{4 a_{0}^4\left(\sqrt{A_{2}(a_{0})}-\sqrt{A_{1}(a_{0})}\right)^2}.
\label{potential}
\end{equation}
It is not difficult to see that it satisfies $V(a_0)=0$ and $V'(a_0)=0$, while the second derivative evaluated at $a_0$ takes the form 
\begin{eqnarray}
V''(a_0)&=& -\frac{5 \left(\sqrt{A_{2}(a_{0})}-\sqrt{A_{1}(a_{0}})\right)^{2}}{a_{0}^2}  -\frac{3\left(\sqrt{A_{1}(a_{0})}+\sqrt{A_{2}(a_{0}})\right)^{2}}{a_{0}^2}\nonumber \\
&&-\frac{\left(A_{2}'(a_{0})-A_{1}'(a_{0})\right)^2}{2\left(\sqrt{A_{2}(a_{0})}-\sqrt{A_{1}(a_{0})}\right)^{2}}
-\frac{4\left(\sqrt{A_{1}(a_{0})}+\sqrt{A_{2}(a_{0}})\right)^{2}\left(A_{2}'(a_{0})-A_{1}'(a_{0})\right)}{a_{0}\left(A_2(a_0)-A_1(a_0)\right)}  \nonumber \\
&&+\frac{A_{1}''(a_{0})+A_{2}''(a_{0})}{2}
-\frac{\left(\sqrt{A_{1}(a_{0})}+\sqrt{A_{2}(a_{0})}\right)^{2}\left(A_{2}''(a_{0})-A_{1}''(a_{0})\right)}{2\left(A_{2}(a_{0})-A_{1}(a_{0})\right)}. 
\label{potencial2der}
\end{eqnarray}
This second derivative allows us to determine when a configuration with radius  $a_0$  is stable: it happens if and only if $V''(a_0)>0$.

\section{Thin-shell wormholes}\label{whs}

For the construction of thin-shell wormholes, we follow the formalism explained in Sect. 2, with some modifications. We start with two spherically symmetric spacetimes of the form (\ref{sphericalmetric}) from which we cut the outer manifolds $\mathcal{M}_{1,2 }=\{X^{\alpha }=(t_{1,2},r,\theta,\varphi)/r\geq a\}$ and paste them on the hypersurface $\Sigma$ determined by $G(r)=r-a=0$, so we obtain a new manifold $\mathcal{M}=\mathcal{M}_{1} \cup \mathcal{M}_{2}$. This construction represents a wormhole compounded by two different regions connected by a throat with radius $a$, where the flare-out condition is satisfied, because the area $4\pi r^2$ is minimal at $r=a$.  We can define a global radial coordinate $\ell \in \mathbb{R}$ in terms of the proper radial distance taken from each side to the throat $\ell =\pm \int_{a}^{r}\sqrt{1/A_{1,2} (r)}dr$, with the ($\pm$) signs that refer, respectively, to $\mathcal{M}_{1,2}$. Again, the angular coordinates are mutually identified and the proper time at both sides of  $\Sigma$ should be the same. The whole manifold $\mathcal{M}$ has coordinates $X^{\alpha }_{1,2} = (t_{1,2},r,\theta, \varphi)$, while at $\Sigma$ we adopt coordinates $\xi ^{i}=(\tau ,\theta,\varphi )$, with $\tau $ the proper time; we let the throat radius depend on $\tau$. In the previously defined orthonormal basis, the first fundamental form at the shell reads $h^{1,2}_{\hat{\imath}\hat{\jmath}}= \mathrm{diag}(-1,1,1)$. The unit normals at $\Sigma$ are given by Eq. (\ref{general_normal_fr}), but with the one related to $\mathcal{M}_{2}$ now inverted, so that 
\begin{equation}
\label{WH_normal}
n_{\gamma }^{1,2}= \pm  \left(-\dot{a},\frac{\sqrt{A_{1,2}(a)+\dot{a}^2}}{A_{1,2}(a)},0,0 \right),
\end{equation}
where the ($\pm$) signs correspond to the different regions $\mathcal{M}_{1,2}$, respectively. By using Eq. (\ref{sff}), the non-null components of the second fundamental form are
\begin{equation}
\label{WH_Ktemporal}
K_{\hat{\tau}\hat{\tau}}^{1,2}= \pm\frac{A '_{1,2}(a)+2\ddot{a}}{2\sqrt{A_{1,2}(a)+\dot{a}^2}}
\end{equation}
and
\begin{equation}
\label{WH_Kangular}
K_{\hat{\theta}\hat{\theta}}^{1,2}=K_{\hat{\varphi}\hat{\varphi}}^{1,2
}=\mp\frac{1}{a}\sqrt{A_{1,2} (a) +\dot{a}^2}.
\end{equation}
The radius of possible solutions is given by Eq. (\ref{conditionEq}), which in this case results in
\begin{equation}
\label{WH_EqCondition1}
\frac{2a\ddot{a}+a A_{1}'(a) + 4(A_{1}(a)+\dot{a}^2)}{\sqrt{A_{1}(a)+\dot{a}^2}}+\frac{2a\ddot{a}+a A_{2}'(a) + 4(A_{2}(a)+\dot{a}^2)}{\sqrt{A_{2}(a)+\dot{a}^2}}=0,
\end{equation}
which written in another form reads
\begin{equation}
\label{WH_EqCondition2}
\frac{2\ddot{a}+ A_{1}'(a)}{2\sqrt{A_{1}(a)+\dot{a}^2}}+\frac{2\ddot{a}+ A_{2}'(a)}{2\sqrt{A_{2}(a)+\dot{a}^2}}=-\frac{2}{a}\left(\sqrt{A_{1}(a)+\dot{a}^2}+\sqrt{A_{2}(a)+\dot{a}^2}\right).
\end{equation}
In the context of non-quadratic $F(R)$, we additionally have to require Eq. (\ref{R_same}) that once again forces $R_1(a)=R_2(a) \equiv R_\Sigma $.  The field equations at the shell $\Sigma $ are given by Eq. (\ref{fieldeq}), which allows us to obtain the energy density and the pressure
\begin{eqnarray}
\label{WH_energy_gral}
\sigma &=& \frac{ F'(R_\Sigma)}{2\kappa }\left(\frac{2\ddot{a}+A_{1}'(a)}{\sqrt{A_{1}(a)+\dot{a}^2}}+ \frac{2\ddot{a}+A_{2}'(a)}{\sqrt{A_{2}(a)+\dot{a}^2}}\right) \nonumber \\
&& +\frac{ F''(R_\Sigma)}{\kappa }\left( R_1'(a)\sqrt{A_{1}(a)+\dot{a}^2}+ R_2'(a)\sqrt{A_{2}(a)+\dot{a}^2}\right),
\end{eqnarray}
and 
\begin{eqnarray}
\label{WH_pressure_gral}
p &=& -\frac{F'(R_\Sigma)}{\kappa a}\left( \sqrt{A_{1}(a)+\dot{a}^2}+ \sqrt{A_{2}(a)+\dot{a}^2}\right) \nonumber \\
&& -\frac{F''(R_\Sigma)}{\kappa }\left( R_1'(a)\sqrt{A_{1}(a)+\dot{a}^2}+  R_2'(a)\sqrt{A_{2}(a)+\dot{a}^2}\right),
\end{eqnarray}
respectively. 
Using Eq. (\ref{WH_EqCondition2}), we can rewrite the energy density to give
\begin{eqnarray}
\label{WH_sigma_rewriten}
\sigma &=& -\frac{2 F'(R_\Sigma)}{\kappa a}\left( \sqrt{A_{1}(a)+\dot{a}^2}+ \sqrt{A_{2}(a)+ \dot{a}^2}\right) \nonumber \\
&& +\frac{F''(R_\Sigma)}{\kappa }\left( R_1'(a)\sqrt{A_{1}(a)+\dot{a}^2}+  R_2'(a)\sqrt{A_{2}(a)+\dot{a}^2}\right).
\end{eqnarray}
It is easy to see that the energy density and the pressure satisfy the equation of state (\ref{state_eq:dynamic}) where, in this case, the brane tension is now given by
\begin{eqnarray}
\label{WH_brane_tension}
\lambda=-\frac{F''(R_\Sigma)}{\kappa }\left( R_1'(a)\sqrt{A_{1}(a)+\dot{a}^2}+  R_2'(a)\sqrt{A_{2}(a)+\dot{a}^2}\right).
\end{eqnarray}
To obtain the radius $a_0$ of the static solutions, we need to solve Eq. (\ref{WH_EqCondition1}), which reduces to
\begin{equation}
\label{WH_static_eqcond1}
\frac{a_0 A_{1}'(a_0) + 4A_{1}(a_0)}{\sqrt{A_{1}(a_0)}}+\frac{a_0 A_{2}'(a_0)+ 4A_{2}(a_0)}{\sqrt{A_{2}(a_0)}}=0,
\end{equation}
or rewritten in another useful way
\begin{equation}
\label{WH_static_eqcond2}
\frac{A_{1}'(a_0)}{2\sqrt{A_{1}(a_0)}}+\frac{A_{2}'(a_0)}{2\sqrt{A_{2}(a_0)}}=-\frac{2}{a_0}\left(\sqrt{A_{1}(a_0)}+\sqrt{A_{2}(a_0)}\right).
\end{equation}
The energy density and the pressure at $\Sigma$ in the static case are
\begin{equation}
\label{WH_static_sigma}
\sigma_0 = \frac{F'(R_\Sigma)}{2\kappa }\left( \frac{A_{1}'(a_0)}{\sqrt{A_{1}(a_0)}}+ \frac{A_{2}'(a_0)}{\sqrt{A_{2}(a_0)}}\right)+ \frac{F''(R_\Sigma)}{\kappa }\left( R_1'(a_0)\sqrt{A_{1}(a_0)}+ R_2'(a_0)\sqrt{A_{2}(a_0)}\right)
\end{equation}
and
\begin{equation}
\label{WH_static_pressure}
p_0 = -\frac{F'(R_\Sigma)}{\kappa a_0} \left( \sqrt{A_{1}(a_0)}+ \sqrt{A_{2}(a_0)}\right)- 
\frac{F''(R_\Sigma)}{\kappa }\left( R_1'(a_0)\sqrt{A_{1}(a_0)}+ R_2'(a_0)\sqrt{A_{2}(a_0)}\right).
\end{equation}
As above, the density energy can be rewritten in the form
\begin{equation}
\label{WH_static_sigma_with_cond}
\sigma_0 = -\frac{2 F'(R_\Sigma)}{\kappa a_0} \left( \sqrt{A_{1}(a_0)}+ \sqrt{A_{2}(a_0)}\right)+ 
\frac{F''(R_\Sigma)}{\kappa }\left( R_1'(a_0)\sqrt{A_{1}(a_0)}+ R_2'(a_0)\sqrt{A_{2}(a_0)}\right).
\end{equation}
From Eq. (\ref{WH_brane_tension}), the brane tension, appearing in the static case equation of state (\ref{state_eq:static}), reads
\begin{eqnarray}
\label{WH_static_branetension}
\lambda _0=-\frac{F''(R_\Sigma)}{\kappa }\left( R_1'(a_0)\sqrt{A_{1}(a_0)}+ R_2'(a_0)\sqrt{A_{2}(a_0)}\right).
\end{eqnarray}
Similarly to what has been done in Sect. 2, we use $\ddot{a}= (1/2)d(\dot{a}^2)/da$ and the definition of $z=\sqrt{A_{2}(a)+\dot{a}^2}+\sqrt{A_{1}(a)+\dot{a}^2}$, which allow us to rewrite Eq. (\ref{WH_EqCondition1}) as $az'(a)+2z(a)=0$. Its solution takes the form given by Eq. (\ref{condicionPot}), but now the potential is
\begin{equation}
\label{WH_static_potencial}
V(a)= -\frac{a_{0}^4\left(\sqrt{A_{1}(a_{0})}+\sqrt{A_{2}(a_{0})}\right)^2}{4a^4} +\frac{A_{1}(a)+A_{2}(a)}{2} -\frac{a^4 \left(A_{1}(a)-A_{2}(a)\right)^{2}}{4 a_{0}^4\left(\sqrt{A_{1}(a_{0})}+\sqrt{A_{2}(a_{0})}\right)^2},
\end{equation}
which satisfies $V(a_0)=0$ and $ V'(a_0)=0 $, while its second derivative at $a_0$ is
\begin{eqnarray}
V''(a_0)&=& -\frac{5 \left(\sqrt{A_{1}(a_{0})}+\sqrt{A_{2}(a_{0}})\right)^{2}}{a_{0}^2}+\frac{A_{1}''(a_{0})+A_{2}''(a_{0})}{2}\nonumber \\
&& -\frac{3\left( A_{1}(a_{0})-A_{2}(a_{0})\right) ^2}{a_{0}^2\left(\sqrt{A_{1}(a_{0})}+\sqrt{A_{2}(a_{0}})\right)^{2}}
-\frac{4\left(A_{1}(a_{0})-A_{2}(a_{0})\right)\left(A_{1}'(a_{0})-A_{2}'(a_{0})\right)}{a_{0}\left(\sqrt{A_{1}(a_{0})}+\sqrt{A_{2}(a_{0})}\right)^{2}}  \nonumber \\
&& -\frac{\left(A_{1}'(a_{0})-A_{2}'(a_{0})\right)^2}{2\left(\sqrt{A_{1}(a_{0})}+\sqrt{A_{2}(a_{0})}\right)^{2}}-\frac{\left(A_{1}(a_{0})-A_{2}(a_{0})\right)\left(A_{1}''(a_{0})-A_{2}''(a_{0})\right)}{2\left(\sqrt{A_{1}(a_{0})}+\sqrt{A_{2}(a_{0})}\right)^{2}}.
\label{WH_static_potencialder2}
\end{eqnarray}
The static solution with throat radius $a_0$ is stable only if it obeys $ V''(a_0)>0$.

\section{Examples}\label{examples}

In order to provide concrete examples, we start with the action corresponding to the theory of gravity defined by $F(R)= R+2\beta\sqrt{R-8\Lambda}-\Lambda$ and coupled to a Maxwell field
\begin{equation}
S=\frac{1}{2 \kappa}\int d^4x \sqrt{|g|} (F(R)-\mathcal{F}_{\mu\nu}\mathcal{F}^{\mu\nu}),
\label{action} 
\end{equation} 
where $g=\det (g_{\mu \nu})$ is the determinant of the metric, $\mathcal{F}_{\mu \nu }=\partial _{\mu }\mathcal{A}_{\nu } -\partial _{\nu }\mathcal{A}_{\mu }$ is the electromagnetic tensor, $\beta$ is a parameter, and $\Lambda$ is twice the conventional cosmological constant. From this action, the field equations for a non-constant scalar curvature $R(r)$ admit a spherically symmetric solution of the form given by Eq. (\ref{sphericalmetric}), in which the metric function reads \cite{rnonconst4}
\begin{equation} 
A(r) = \frac{1}{2}-\frac{M}{r}+\frac{Q^2}{ r^2}-\frac{2 \Lambda r^2}{3},
\label{A_metric_nonconstant}
\end{equation}
where the mass parameter\footnote{We adopt this name due to the formal resemblance with the usual mass, but it is not a free constant in the solution. } $M=-1/(3 \beta)$ is determined by the dimensional parameter $\beta$ of the theory and the squared charge is $Q^2=Q_\mathrm{E}^2+Q_\mathrm{M}^2$, with $Q_\mathrm{E}$ the electric charge and $Q_\mathrm{M}$ the magnetic charge. The parameter $\beta$ cannot vanish and is always negative (therefore $M$ is positive). A  spacetime with the same form was previously found for a similar Lagrangian coupled to the electromagnetic field \cite{rnonconst2}, but  with a negative mass parameter; in this case, it can be interpreted as a global monopole solution in the presence of the Maxwell field\footnote{In the same work \cite{rnonconst2}, the study is also extended to a particular case of power-law nonlinear electrodynamics for which a black hole solution is obtained.}. Using Eq. (\ref{Rnonconstant}), the scalar curvature for this geometry is given by
\begin{equation}
\label{Rnonconstant_equation}
R(r)=\frac{1}{r^2}+8\Lambda. 
\end{equation}
Then, the expressions of $F(R)$, $F'(R)$, and $F''(R)$ can be written as functions of $r$ in terms of $\Lambda$ and $\beta$, resulting in
\begin{equation}
\label{fr_formas1}
F(R) = \frac{1}{r^2}+\frac{2\beta}{r}+7\Lambda,
\end{equation}
\begin{equation}
\label{fr_formas2}
F'(R)= 1+r\beta,
\end{equation}
and
\begin{equation}
\label{fr_formas3}
F''(R)= -\frac{\beta}{2}r^3 .
\end{equation}
Let us review the horizon structure of the solution. The possible radii of the horizons are given by the zeros of the metric function $A(r)$. If $\Lambda = 0$ and $Q=0$, there is only one (Schwarzschild like) event horizon with radius $r_h=2M$. When $\Lambda = 0$ and $Q \neq 0$, there are two (Reissner-Nordstr\"om like) horizons: the inner one with $r_i=M-\sqrt{M^2-2Q^2}$ and the (outer) event one with $r_h=M+\sqrt{M^2-2Q^2}$, resulting from solving a quadratic equation. The geometry in this case also presents a critical charge $Q_c=M/\sqrt{2}$, corresponding to the value of charge for which the inner horizon and the event horizon fuse into one; beyond $ | Q |> Q_c $, there only exist a naked singularity. When $\Lambda \neq 0$ and $Q = 0$, for $\Lambda < 0$  there is only the event horizon ($r_h$), while for $0< \Lambda< 1/(36M^2)$ the cosmological horizon ($r_c$) is present in addition to the event horizon ($r_h$); in both cases (Schwarzschild anti de Sitter/de Sitter like, respectively), they are obtained as the real and positive solutions of a cubic equation\footnote{The analytic expressions for the radii of the horizons are cumbersome, so they are not shown here.\label{cumbersome}}. When $\Lambda \neq 0$ and $Q \neq 0$ (Reissner-Nordstr\"om  anti de Sitter/ de Sitter like), the horizons are given by the roots of a fourth-degree polynomial\footnote{The remark of F.N. \ref{cumbersome} applies again.}. In this case, a critical value of the charge $Q_c $ also exists. For $\Lambda<0$, there is an inner ($r_i$) and an event horizon ($r_h$) if $0 < | Q | <Q_c $, while if $ | Q |> Q_c $ there appears a naked singularity. For $0< \Lambda <\Lambda_\mathrm{max}$,  when $0 < | Q |< Q_c $ there is a cosmological horizon ($r_c$) in addition to the event ($r_h$) and the inner ($r_i$) horizons, while when $ | Q |> Q_c $ the cosmological horizon coexists with a naked singularity. For a given $M$, the value of $\Lambda_\mathrm{max} \geq  1/(36 M^2)$ increases with $|Q|$.

\subsection{Thin shell with charge surrounding a black hole}

With the purpose of constructing a manifold $\mathcal{M}$ having a thin shell of matter at $\Sigma$, we adopt the interior region $\mathcal{M}_1 $ as the one described by the metric function in Eq. (\ref{A_metric_nonconstant}) with a null charge $Q_1=0$,  
\begin{equation} 
A_1 (r) = \frac{1}{2}-\frac{M}{r}-\frac{2 \Lambda r^2}{3},
\label{A_metric_nonconstant_interior}
\end{equation}
while we take the exterior region $\mathcal{M}_2 $ with $Q_2=Q\neq 0$ 
\begin{equation} 
A_2 (r) = \frac{1}{2}-\frac{M}{r}+\frac{Q^2}{ r^2}-\frac{2 \Lambda r^2}{3}.
\label{A_metric_nonconstant_exterior}
\end{equation}
The mass parameter $M$ is the same for both regions since $M=-1/ (3 \beta) $. The inner geometry corresponds to a black hole, for $\Lambda \leq 0$ with only the event horizon with radius $r^{(1)}_h$, while for $0<\Lambda<1/(36M^2)$ besides it there is a cosmological horizon with radius $r^{(1)}_c$. The radius of the surface $\Sigma$ is taken so that $r^{(1)}_h < a$ in the first case and $r^{(1)}_h < a <r^{(1)}_c$ in the second one. The outer geometry has an event horizon radius $r^{(2)}_h $ that always satisfies  $r^{(2)}_h < r^{(1)}_h $ for any value of the charge, so our construction removes it; in the case with $0<\Lambda<1/(36M^2)$, the condition $a < r^{(2)}_c$ is also demanded. In this way, the thin shell surrounds the event horizon of a black hole, with no other horizons if $\Lambda \leq 0$, or with a cosmological horizon having a radius larger than $a$ if $\Lambda >0$. For a proper matching at the shell, the fulfillment of Eqs. (\ref{conditionEq}) and (\ref{R_same})  is required. The first one leads the radius $a$ to obey Eq. (\ref{KJump:dynamic01}). It is easy to verify that $R_1(a)=R_2(a)$, so the second one is automatically satisfied. At the shell, the energy density is given by Eq. (\ref{sigma:dynamic_plus_cond}) and the pressure by Eq. (\ref{pressure:dynamic}).
 
In order to proceed with the study of the existence and the stability of static shells, we take a constant radius $a_0$. As explained above, the value of  $a_0$ is larger than the event horizon radius of $\mathcal{M}_1$ and, in the case of $\Lambda>0$, also smaller than the cosmological horizon radius of $\mathcal{M}_2$; our construction automatically removes the event horizon of $\mathcal{M}_2$ and the cosmological horizon of $\mathcal{M}_1$ when $\Lambda>0$. The whole manifold $\mathcal{M}$ then consists of a black hole without charge surrounded by a static charged thin shell. After checking that $R_1(a_0)=R_2(a_0)$ so that Eq. (\ref{R_same}) is fulfilled, we demand the radius $a_0$ to satisfy Eq. (\ref{KJump:static01}). If we want to guarantee that the shell is made of normal matter, we require the fulfillment of the weak energy condition. By using Eqs. (\ref{fr_formas2}) and (\ref{fr_formas3}), the energy density and the pressure at $\Sigma$ can be obtained from Eqs. (\ref{sigma:static_plus_cond}) and (\ref{pressure:static_plus_cond}), which for this case have the form
\begin{equation}
 \sigma_0 = -\frac{ 1}{ \kappa }\left( \frac{2}{a_0}+3 \beta  \right)  \left( \sqrt{A_{2}(a_0)}-\sqrt{A_{1}(a_0)} \right) 
 \label{sigma_pegado_final}
\end{equation}
and
\begin{equation}
p_0 = -\frac{ 1}{\kappa a_0} \left( \sqrt{A_{2}(a_0)}-\sqrt{A_{1}(a_0)} \right) .
 \label{press_pegado_final}
\end{equation} 
They satisfy the equation of state given by Eq. (\ref{state_eq:static}), with the brane tension
\begin{equation}
\lambda _0 = \frac{ \beta}{\kappa } \left( \sqrt{A_{2}(a_0)}-\sqrt{A_{1}(a_0)} \right) .
 \label{tension_pegado_final}
\end{equation}
To analyze the weak energy condition, given by the inequalities $\sigma_0 \geq 0$ and $\sigma_0 + p_0 \geq 0$, we use Eqs. (\ref{sigma_pegado_final}) and (\ref{press_pegado_final}), obtaining the following expression
\begin{equation}
\sigma_0 + p_0 = -\frac{ 3}{\kappa } \left( \frac{1}{a_0}+\beta  \right) \left( \sqrt{A_{2}(a_0)}-\sqrt{A_{1}(a_0)} \right) .
 \label{weakenergycond}
\end{equation} 
To guarantee that $\sigma_0\geq 0$, we observe from Eq. (\ref{sigma_pegado_final}) that its sign is given by the sign of the two factors within brackets. The last factor, which is the difference of the square roots of the metric functions, is always positive for any value of charge. Therefore, the overall sign of $\sigma_0$ is given by the other factor, which has to be negative, so it leads to 
\begin{equation}
a_0 \geq \frac{2}{3|\beta |}=2M.
\label{weakenergycond1}
\end{equation}
By doing a similar analysis for Eq. (\ref{weakenergycond}), we obtain that $\sigma_0 + p_0 \geq 0$ gives
\begin{equation}
a_0 \geq \frac{1}{|\beta |} =3M.
\label{weakenergycond2}
\end{equation}
The weak energy condition is satisfied (normal matter) when Eqs. (\ref{weakenergycond1}) and (\ref{weakenergycond2}) are both fulfilled, so it reduces to the inequality in Eq. (\ref{weakenergycond2}). As mentioned in Sect. \ref{tsh}, it is desirable to add the condition $F'(R)=1+a_0\beta>0$ in order to avoid the presence of ghosts at the shell, leading to 
\begin{equation}
a_0< \frac{1}{|\beta |}=3M,
\label{condsobrebeta}
\end{equation}
which by comparing with Eq. (\ref{weakenergycond2}) means that the zone of normal matter coincides with the zone of ghosts.
\begin{figure}[t!]
\centering
\includegraphics[width=0.9\textwidth]{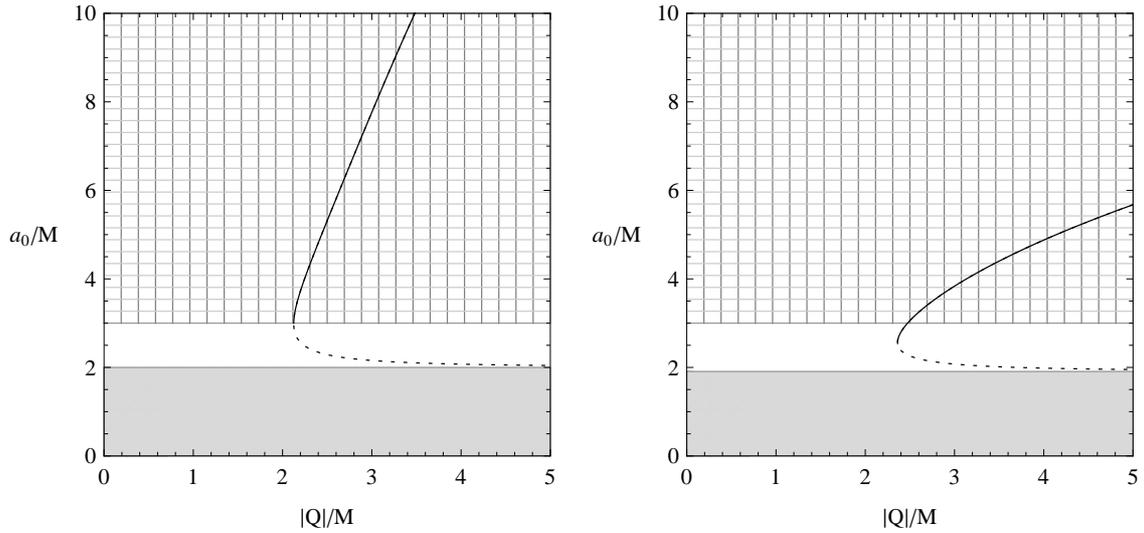}
\caption{Thin shells surrounding a black hole in the theory $F(R)=R+2\beta \sqrt{R-8\Lambda} -\Lambda$, for negative values of $\Lambda$. The solid lines represent the stable static solutions with radius $ a_0 $, while the dotted lines the unstable ones. The mass parameter is $M=-1/(3\beta)$ in both regions separated by the shell and the charge $ Q $ corresponds to the exterior geometry. The meshed zones represent normal matter, while the gray areas have no physical meaning (see text). Left: $\Lambda M^2=-1 \times 10^{-4}$; right: $\Lambda M^2=-1 \times 10^{-2} $. }
\label{figLneg}
\end{figure}
\begin{figure}[t!]
\centering
\includegraphics[width=0.9\textwidth]{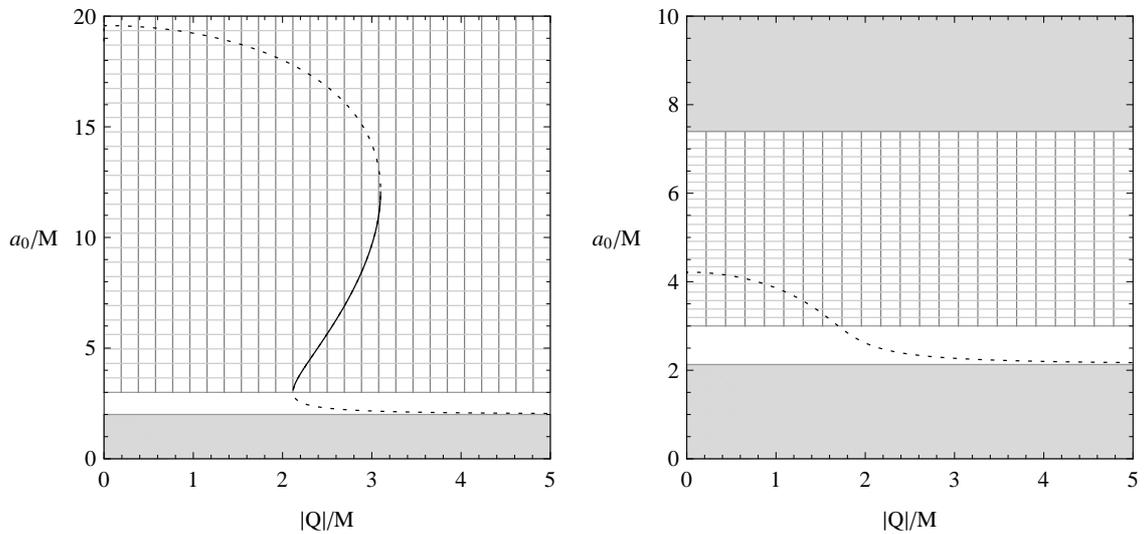}
\caption{Same as Fig. \ref{figLneg} but for positive values of $\Lambda$. Left: $\Lambda M^2=1 \times 10^{-4}$, in this case there is another non-physical region, not shown, that is located at $a_0/M \gtrsim 85$; right: $\Lambda M^2=1 \times 10^{-2} $.}
\label{figLpos} 
\end{figure}

For the stability analysis of a static solution with radius $a_0$, we use Eq. (\ref{potencial2der}), which assures that it is stable if and only if $V''(a_0)>0$. The results are shown in Figs. \ref{figLneg} and \ref{figLpos}, which were selected as the most representative ones. The meshed zones represent those where the solution satisfies the weak energy condition, while the gray zones lack physical meaning. Solid lines correspond to stable solutions; the unstable ones are displayed with dotted lines. The basic behavior of the solutions does not change with the different values of $M$ (or equivalently of $\beta $), only its scale. Hence, the quantities presented in them are adimensionalized with the mass parameter $M$. The main features are
\begin{itemize}
\item For $\Lambda<0$, we find that the static solutions only exist from a certain value of $|Q|/M>2$. From that value on, we find two solutions, one of them unstable and close to the event horizon, while the other is stable and with $a_0/M$ monotonously increasing with $|Q|/M$. The unstable solution is made of exotic matter, while the stable one is constituted by normal matter in most for the range of $a_0/M$. 
\item For $\Lambda>0$ but close to zero, there exist unstable solutions for small values of $|Q|/M$. There is a short range of values of $|Q|/M$ for which there exist three solutions, two of them unstable and one stable. Only one of the unstable solutions is made of normal matter, while the other one is constituted by exotic matter. The stable solution is always made of normal matter. For large values of $|Q|/M$, there exists only one unstable solution close to the event horizon with exotic matter. 
\item For  $0<\Lambda<1/(36M^2)$ but close to $1/(36M^2)$, there is only one unstable solution that gets closer to the event horizon for large values of $|Q|/M$. For small values of $|Q|/M$, it is constituted by normal matter. 
\end{itemize}

\subsection{Charged thin-shell wormhole}

Now, we proceed to construct a traversable thin-shell wormhole with identical gravitational Lagrangians in the regions at the sides of the throat, i.e., the same $F(R)$ theory, and allowing different electromagnetic fields in them. Thus, we take two  copies $\mathcal{M}_{1,2} $ of the outer region ($ r\geq a$) of the geometry defined by Eq. (\ref{A_metric_nonconstant}), both having the same $\Lambda$ and mass parameter $M=-1/ (3 \beta)$, but with values of charge $Q_1$ and $Q_2$, respectively, that is
\begin{equation} 
A_1 (r) = \frac{1}{2}-\frac{M}{r}+\frac{Q_1^2}{ r^2}-\frac{2 \Lambda r^2}{3},
\label{A_metric_nonconstant_wh1}
\end{equation}
and
\begin{equation} 
A_2 (r) = \frac{1}{2}-\frac{M}{r}+\frac{Q_2^2}{ r^2}-\frac{2 \Lambda r^2}{3}.
\label{A_metric_nonconstant_wh2}
\end{equation}
We paste them at the spherical surface $\Sigma$, with radius $a$, to give $\mathcal{M}=\mathcal{M}_1 \cup \mathcal{M}_2$. The value of $a$ is chosen larger than the radii of both event horizons and smaller than the radii of the cosmological horizons, when any of them exist. This construction guarantees the traversability and the flare-out condition since the area $4\pi a^2$ at $\Sigma$ is minimal, so this surface corresponds to the throat. The radius $a$ has to fulfill Eq. (\ref{WH_EqCondition1}), while $\sigma$ and $p$ are given by Eqs. (\ref{WH_sigma_rewriten}) and (\ref{WH_pressure_gral}), respectively, and they are related to the brane tension shown in Eq. (\ref{WH_brane_tension}) by the equation of state displayed in Eq.  (\ref{state_eq:dynamic}).

For the static case, with a throat radius $a_0$, the condition given by Eq. (\ref{WH_static_eqcond1}) should be fulfilled. The expressions for the energy density and the pressure at $\Sigma$, given by Eqs. (\ref{WH_static_pressure}) and (\ref{WH_static_sigma_with_cond}), by using Eqs. (\ref{fr_formas2}) and (\ref{fr_formas3}) can be written in the form 
\begin{equation}
\label{WH_static_asymetric_energy}
\sigma _0= -\frac{1}{\kappa}\left( \frac{2}{a_0}+ \beta \right) \left( \sqrt{A_1(a_0)}+ \sqrt{A_2(a_0)} \right)
\end{equation}
and
\begin{equation}
\label{WH_static_asymetric_presure}
p_0=-\frac{1}{\kappa}\left( \frac{1}{a_0}+ 2\beta \right)\left( \sqrt{A_1(a_0)}+ \sqrt{A_2(a_0)} \right).
\end{equation}
The equation of state has the form shown in Eq. (\ref{state_eq:static}), where the brane tension is now given by
\begin{equation}
\label{WH_static_asymetric_brane_tension}
\lambda_0=-\frac{2 \beta}{\kappa}\left( \sqrt{A_1(a_0)}+ \sqrt{A_2(a_0)} \right).
\end{equation}
Taking into account these equations, we can analyze the weak energy condition. From $\sigma_0 \geq0$ we obtain the inequality
\begin{equation}
\label{WH_weak_energ_sigma}
a_0\geq\frac{2}{|\beta|}=6M,
\end{equation}
while the sum
\begin{equation}
\label{WH_weak_energ_sigm_press}
\sigma_0+p_0=-\frac{3}{\kappa}\left( \frac{1}{a_0}+ \beta \right)\left( \sqrt{A_1(a_0)}+ \sqrt{A_2(a_0)} \right)
\end{equation}
is non-negative only if Eq. (\ref{weakenergycond2}) is satisfied. The weak energy condition then requires that Eq. (\ref{WH_weak_energ_sigma}) should be fulfilled, since it is the more restrictive of them. It is worth remembering that the condition $F'(R(a_0))>0$ is equivalent to the inequality in Eq. (\ref{condsobrebeta}), so once again we have the presence of ghosts when the matter at the shell is normal. 
In order to analyze the stability, we employ the potential given by Eq. (\ref{WH_static_potencial}), satisfying $V(a_0)=0$ and $V'(a_0)=0$; therefore, we use Eq. (\ref{WH_static_potencialder2}) to determine that a static solution with radius $a_0$ is stable when $V''(a_0)>0$. 
\begin{figure}[t!]
\centering
\includegraphics[width=0.9\textwidth]{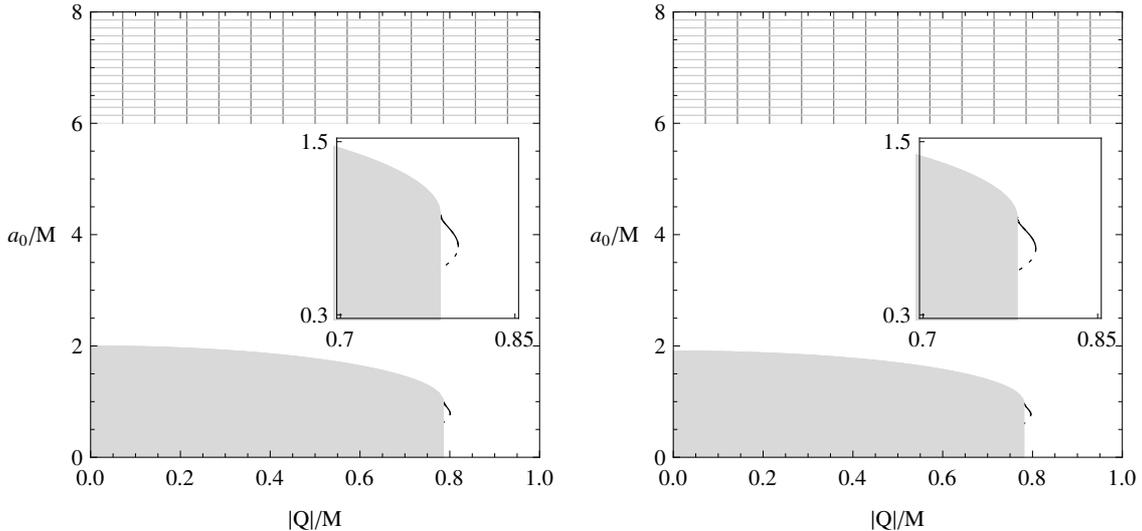}
\caption{Thin-shell wormholes in the theory $F(R)=R+2\beta \sqrt{R-8\Lambda} -\Lambda$, for negative values of $\Lambda$. The solid lines represent the stable static solutions with radius $ a_0 $, while the dotted lines the unstable ones. The mass parameter is $M=-1/(3\beta)$, while the values of charge corresponding to the regions separated by the throat are $Q_1=Q$ and $Q_2= \pm 0.9 Q$. The meshed zones represent normal matter, while the gray areas have no physical meaning (see text). Left: $\Lambda M^2=-1 \times 10^{-4}$; right: $\Lambda M^2=-1 \times 10^{-2} $. }
\label{whfigLneg}
\end{figure}
\begin{figure}[t!]
\centering
\includegraphics[width=0.9\textwidth]{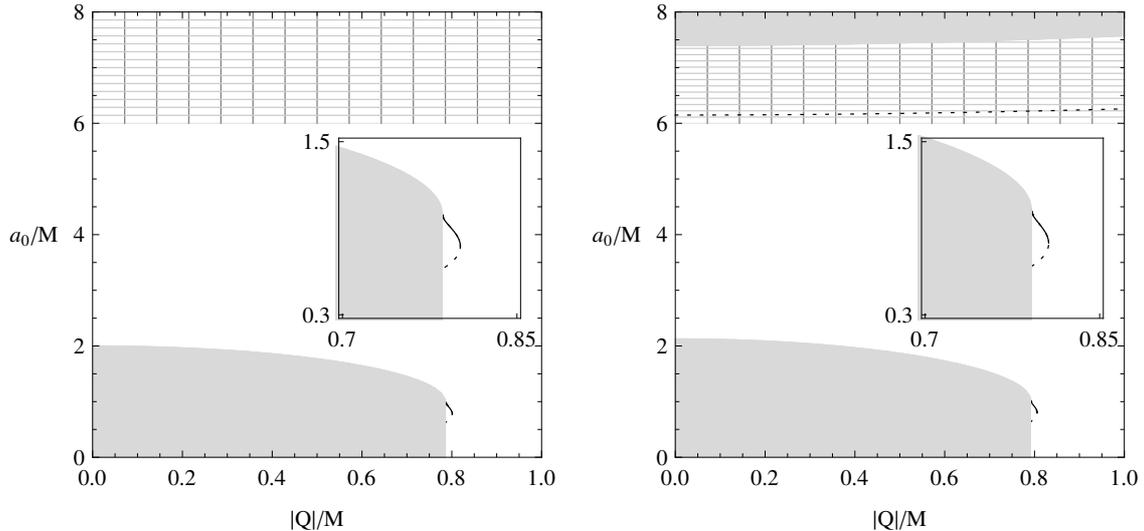}
\caption{Same as Fig. \ref{whfigLneg} but for positive values of $\Lambda$. Left: $\Lambda M^2=1 \times 10^{-4}$, in this case there is another unstable solution, not shown, made of normal matter for $a_0/M \approx 70$, and also a non-physical region, not shown, that is located at $a_0/M \gtrsim 85$. Right: $\Lambda M^2=1 \times 10^{-2} $.}
\label{whfigLpos} 
\end{figure}

We adopt in what follows for the values of charge associated with the regions separated by the throat the notation $Q_1=Q$ and $Q_2=\zeta Q$, with $0 \le |\zeta | \le 1$. Some results are shown in Figs. \ref{whfigLneg} and \ref{whfigLpos}, in which $\zeta =\pm 0.9$. The basic behavior of the solutions does not change with the different values of $M$ (or equivalently of $\beta $); only its scale varies, so all quantities are adimensionalized with the mass parameter. The main characteristics of the solutions are
\begin{itemize}
\item For $\Lambda<0$ and only for a short range of values of $|Q|/M$, larger than the critical charge $Q_c/M$, there exist two solutions. One of them is stable and the other unstable, both made of exotic matter and without the presence of ghosts. 
\item For $\Lambda>0$, two solutions, similar to those found for $\Lambda<0$, coexist with a third one which is present for all values of $|Q|/M$, even without charge. This last solution, close to the non-physical meaning zone located at large values of $a_0/M$, is unstable and made of normal matter in the presence of ghosts.
\end{itemize}
For any sign of $\Lambda$, increasing $|\zeta |$ from the value used in the plots extends the range of $|Q|/M$ for which is present the pair of solutions that includes the stable configuration, while decreasing it reduces this range until this pair of solutions finally disappears. In particular, the largest range of $|Q|/M$ for stability is obtained when $|\zeta |=1$, which corresponds to a thin-shell  wormhole having a geometry symmetric across the throat. Regarding this last case, a related study has been recently published \cite{whfrtsnew} within the context of a special case of a power-law nonlinear electrodynamics, which includes the Maxwell limit.

\section{Discussion}\label{discu}

In this article, we have studied two wide classes of spherically symmetric manifolds with a thin shell in non-quadratic $F(R)$ gravity, with non-constant scalar curvature $R$. One corresponds to thin shells joining an inner geometry with an outer one and the other to thin-shell wormholes. In both cases, we have found general expressions for the energy density $\sigma$ and the isotropic pressure $p$ of the matter at the shell, and we have shown that they satisfy the equation of state $\sigma -2p=-3\lambda$, relating them to the brane tension $\lambda$. We have also analyzed the stability of the static solutions under perturbations that preserve the symmetry. We have provided examples in which each general formalism is applied to a black hole solution of the theory defined by the gravitational Lagrangian $F(R)=R+2\beta\sqrt{R-8\Lambda}-\Lambda$, coupled to the Maxwell electromagnetic field.

In the example corresponding to the first class, the resulting spacetime can be interpreted as a non-charged black hole surrounded by a thin shell of matter with charge $Q$. We have seen that different values of $\beta $ do not modify the qualitative behavior of the solutions but only their scales, so we have adimensionalized all quantities with the mass parameter $M= -1/ (3 \beta) >0$. For negative $\Lambda$ and from a certain value of $|Q|/M$ on, we have found two solutions; one unstable close to the event horizon and another stable which is constituted by normal matter for large enough values of $|Q|/M $. For positive $\Lambda$ but close to zero, we find a range of values of $|Q|/M$ where there exist three solutions; two of them unstable and the other stable, this last one is constituted of normal matter. There are also unstable solutions consisting of normal matter for small values of $|Q|/M$. For positive values of $\Lambda$ and close to $1/(36M^2)$, we have found an unstable solution for all values of $|Q|/M$, with normal matter for small enough values of  $|Q|/M$. By doing the analysis of the weak energy condition at the shell, we have found that the zone of normal matter coincides with the zone where there exist ghost fields, which is not a desirable aspect of the model. Finally, we can say that for suitable values of the parameters, stable solutions are possible, which can have normal matter in the presence of ghosts or exotic matter in the absence of ghosts. 

In the example of the second class, we have constructed a thin-shell wormhole with values of charge $Q_1=Q$ and $Q_2= \zeta Q$, with $0\le |\zeta| \le 1$, at the regions joined by the throat. We have taken the shell radius large enough to prevent the presence of the event horizons, so that the wormhole is traversable. As in the case above, the value of $\beta$ (or $M$) only affects the scale of the results and we use the same adimensionalization. For both negative and positive $\Lambda$, when $\zeta$ is large enough we have found two solutions from values of $|Q|/M$ larger than the critical one, for a short range of  $|Q|/M$. One of them is stable and the other is unstable, both composed of exotic matter. The largest range of $|Q|/M$ for the presence of this pair is obtained for a wormhole geometry which is symmetric across the throat, i.e., when $|\zeta |=1$. For the particular case of $\Lambda>0$, for any $\zeta$ there exists a third unstable solution made of normal matter for any value of $|Q|/M$, including the chargeless case; this solution is always close to the non-physical region located at a large radius. Through the analysis of the weak energy condition at the throat, we have found that the zone of normal matter is included within the zone where ghost fields are present. Summarizing, for a certain set of parameters of the model, we have found stable solutions made of exotic matter close to the critical charge without the presence of ghosts. But the presence of exotic matter is not troublesome, since it is the usual feature in the case of wormholes.

\section*{Acknowledgments}

This work has been supported by CONICET and Universidad de Buenos Aires.

\end{document}